\documentclass[twocolumn,preprintnumbers]{revtex4}

\usepackage{graphicx}
\usepackage{epsf}
\usepackage{amsmath,amssymb}
\usepackage{color}
\newcommand{\bvec}{\boldsymbol}

\newcommand{\Oa}{^{16}\textrm{O}+\alpha}
\newcommand{\Caa}{^{12}\textrm{C}+2\alpha}
\newcommand{\CBe}{^{12}\textrm{C}+^8\textrm{Be}}
\newcommand{\Ne}{^{20}\textrm{Ne}}

\usepackage{ulem}

\begin{document}
\preprint{KUNS-2801, NITEP 65}
\title{Properties of $K^\pi=0^+_1$,  $K^\pi=2^-$, and $K^\pi=0^-_1$ bands of $^{20}$Ne 
probed via proton and alpha inelastic scattering}

\author{Yoshiko Kanada-En'yo}
\affiliation{Department of Physics, Kyoto University, Kyoto 606-8502, Japan}
\author{Kazuyuki Ogata} 
\affiliation{Research Center for Nuclear Physics (RCNP), Osaka University,
  Ibaraki 567-0047, Japan}
\affiliation{Department of Physics, Osaka City University, Osaka 558-8585,
  Japan}
\affiliation{
Nambu Yoichiro Institute of Theoretical and Experimental Physics (NITEP),
   Osaka City University, Osaka 558-8585, Japan}

\begin{abstract}
The $K^\pi=0^+_1$,  $K^\pi=2^-$, and $K^\pi=0^-_1$ bands of $^{20}$Ne 
are investigated with microscopic structure and reaction calculations 
via proton and $\alpha$ inelastic scattering off $^{20}$Ne.
Structures of $^{20}$Ne are calculated 
with variation after parity and total angular momentum projections in the antisymmetrized molecular dynamics(AMD).
The $K^\pi=0^+_1$ and $K^\pi=0^-_1$ bands have 
$^{16}\textrm{O}+\alpha$ cluster structures, whereas
the $K^\pi=2^-$ band shows a $^{12}\textrm{C}+2\alpha$-like structure.
Microscopic coupled-channel calculations of  proton 
and $\alpha$ scattering off  $^{20}$Ne are
performed by using the proton-nucleus and $\alpha$-nucleus potentials, which are
derived by folding the Melbourne $g$-matrix $NN$ interaction with 
AMD densities of $^{20}$Ne. 
The calculation reasonably reproduces the observed cross sections of 
proton scattering at $E_p=25$--35 MeV and $\alpha$ scattering at $E_\alpha=104$--386 MeV.
Transition properties from the ground to excited states are discussed by reaction analyses of 
proton and $\alpha$ inelastic processes.
Mixing of the  $K^\pi=2^-$ and  $K^\pi=0^-_1$ bands is discussed by detailed analysis of 
the $0^+_1\to 3^-_1$ and  $0^+_1\to 3^-_2$ transitions. 
For the $3^-_1$ state,  mixing of the $K^\pi=0^-_1$ cluster component in the $K^\pi=2^-$ band plays 
an important role in the transition properties. 
\end{abstract}

\maketitle

\section{Introduction}
Cluster structure plays an important role 
in nuclei, in particular, in light-mass regions. A typical cluster structure in $sd$-shell nuclei is 
$\Oa$ cluster structure of $\Ne$ \cite{fujiwara80}. The idea of the $\Oa$ cluster structure has been introduced 
to describe energy levels of the parity-doublet 
$K^\pi=0^+_1$ and $K^\pi=0^-_1$ bands \cite{Horiuchi68}, and extended to describe 
the higher nodal $\Oa$ cluster band assigned as the $K^\pi=0^+_4$
 (labeled as $K^\pi=0^+_\textrm{hn}$) band. 
The structure and $\alpha$-decay properties of $\Oa$ cluster states have been 
intensively 
investigated with potential models and (semi)microscopic cluster models 
\cite{Buck:1995zz,wildermuth72,Nemoto72,Nemoto75,Matsuse73,Matsuse75,Fujiwara79a,Fujiwara79b,Fujiwara79c,fujiwara80,Dufour:1994zz,Zhou:2012zz,Zhou:2013ala}.

In excited states of $\Ne$, further rich phenomena beyond the $\Oa$ cluster structure arise 
from cluster breaking, i.e., internal excitation of $^{16}$O and $\alpha$ clusters. 
For example, in the $K^\pi=0^+_1$ band, the cluster breaking is essential to describe deviation from ideal rotational spectra at the band terminal \cite{KanadaEnyo:1994kw,Kimura:2003uf,Itagaki:2010ha}. 
In addition,  $\Caa$ (or $\CBe$) cluster structure has been considered to discuss the 
$K^\pi=0^+_2$ and $K^\pi=0^+_3$ bands, in the 
frameworks of extended cluster models~\cite{Nemoto75,Fujiwara79a,Fujiwara79b,Fujiwara79c} 
and antisymmetrized molecular dynamics (AMD)~\cite{Taniguchi:2004zz}.
The cluster structures of $\Ne$ have been also discussed 
with mean-field approaches~\cite{Ohta:2004wm,Shinohara:2006jd,Zhou:2015nza,Marevic:2018crl}.
Moreover, the lowest negative-parity band experimentally 
assigned as the $K^\pi=2^-$ band is 
considered to be a particle-hole state or octupole $Y_{32}$ 
vibration~\cite{Hiura72,Fujiwara79c,Kimura:2003uf,Ohta:2004wm,Shinohara:2006jd}. 

It means that two types of negative-parity states
appear in low-energy levels of  $\Ne$, the mean-field type states in the $K^\pi=2^-$ band 
and the $\Oa$ cluster states in the $K^\pi=0^-_1$ band. The former band arises from 
the $K^\pi=2^-$ particle-hole excitation, and the latter is caused by 
the $K^\pi=0^-$ excitation of the inter-cluster motion between $^{16}\textrm{O}$ and $\alpha$ clusters. 
In the experimental levels, the $3^-_1$ (5.62 MeV) and  $3^-_2$ (7.16 MeV) states
are assigned to the $K^\pi=2^-$ and $K^\pi=0^-_1$ bands, respectively. 
Energy levels and in-band $E2$ transitions in each band have been 
reproduced well by theoretical calculations with the 
$(\Oa)+(\CBe)$ coupled-channel orthogonal condition model (CC-OCM) \cite{Fujiwara79c}
and the deformed-basis AMD (def-AMD) \cite{Kimura:2003uf}. However, 
the observed $E3$ transition strength, $B(E3;3^-_1\to 0^+_1)$~\cite{Hausser:1971mml}, for the inter-band transition from 
the $K^\pi=2^-$ band to the ground band is much larger by one order of magnitude than the 
theoretical value of CC-OCM, and inconsistent with the simple interpretation of 
the $K^\pi=2^-$ band as the particle-hole excitation.

In order to investigate structure and transition properties 
of the ground and excited bands, 
electron and hadron scattering experiments 
have been performed for $sd$-shell nuclei. 
For $\Ne$, hadron inelastic scattering such as $(p,p')$ and $(\alpha,\alpha')$ have been investigated 
\cite{Hinterberger:1968odq,Swiniarski72,DeSwiniarski:1973nhi,deSwiniarski:1976rcl,Blanpied:1988ew,Rebel:1972nip,Adachi:2018pql}.
Phenomenological reaction analyses of the $(p,p')$ and $(d,d')$ cross sections 
have suggested again the strong $3^-_1\to 0^+_1$ transition consistently with the 
$B(E3;3^-_1\to 0^+_1)$ data determined by $\gamma$ decays, 
but the $B(E3;3^-_1\to 0^+_1)$ values evaluated from hadron scattering 
show dependence on energies and projectile particles.

In principle, proton and $\alpha$ inelastic scattering can be good probes for transition properties 
from the ground to excited states provided that reliable reaction analyses are available.
Recently, microscopic coupled-channel (MCC) calculations for proton and $\alpha$ scattering 
have been remarkably developed. In the MCC calculations, matter and transition densities of target nuclei 
obtained with microscopic structure models are utilized as inputs of 
coupled-channel reaction calculations in microscopic folding models (MFMs), in which 
nucleon-nucleus and $\alpha$-nucleus potentials are constructed by folding effective $NN$ interactions. 
In our previous studies~\cite{Kanada-Enyo:2019prr,Kanada-Enyo:2019qbp,Kanada-Enyo:2019uvg,Kanada-Enyo:2020zpl}, 
we have applied the MCC calculations to proton and $\alpha$ scattering off
various target nuclei in the $p$- and $sd$-shell regions
with structure model calculation of AMD. Using the Melbourne $g$-matrix 
$NN$ interaction \cite{Amos:2000}, we have succeeded in reproducing 
$(p,p')$ and $(\alpha,\alpha')$ cross sections of various excited states
such as cluster and vibration excitations. 
The Melbourne $g$-matrix interaction is an effective $NN$ interaction in nuclear medium
based on a bare $NN$ interaction of the Bonn B potential~\cite{Mac87}.
Owing to the fundamental derivation, it contains energy and density dependences 
in the applicable range without relying on phenomenological adjustment of interaction parameters. 

In the present study, we calculate structure of $\Ne$ with 
variation after parity and angular-momentum projections (VAP) 
in the AMD framework~\cite{KanadaEnyo:1995tb,KanadaEn'yo:1998rf,KanadaEn'yo:2012bj}. 
We, then, apply the MCC approach to proton and $\alpha$ scattering 
off $\Ne$ with the Melbourne $g$-matrix 
$NN$ interaction using AMD densities of $\Ne$ as structure inputs of the target nucleus.
With analyses of the structure and reaction calculations, 
structures of the ground and excited states in the $K^\pi=0^+_1$, $K^\pi=2^-$, and $0^-_1$ bands 
are investigated. 
In particular, properties of the $3^-_1$ and $3^-_2$ states and possible mixing of the 
$K^\pi=2^-$ and $K^\pi=0^-_1$ bands are discussed in detail. 

The paper is organized as follows. 
The next section describes the frameworks of the 
AMD calculation of $\Ne$ and the MCC approach for proton and $\alpha$ scattering off  $\Ne$. 
Structure properties are described in Sec.~\ref{sec:results1}, 
and the results of proton and $\alpha$ scattering are shown in Sec.~\ref{sec:results2}. 
A discussion of the $3^-_1$ and $3^-_2$ states is given in  Sec.~\ref{sec:discussions}. 
Finally the paper is summarized in Sec.~\ref{sec:summary}. 

\section{Method} \label{sec:method} 


\subsection{Structure calculations of $\Ne$}
We apply the VAP version of AMD to calculate structure of $\Ne$. The method is almost the same as 
those used for studies of $^{12}$C and neutron-rich Be isotopes in 
Refs.~\cite{KanadaEn'yo:1998rf,Kanada-Enyo:1999bsw,Kanada-Enyo:2003fhn}. 
It is sometimes called AMD+VAP, but we simply call it AMD in the present paper. 
For comparison, we also apply a $\Oa$-cluster model with the generator coordinate 
method (GCM)~\cite{GCM1,GCM2}. 

In the framework of AMD,  an $A$-nucleon wave function 
is given by a Slater determinant of 
single-nucleon Gaussian wave functions as
\begin{eqnarray}
 \Phi_{\rm AMD}({\bvec{Z}}) &=& \frac{1}{\sqrt{A!}} {\cal{A}} \{
  \varphi_1,\varphi_2,...,\varphi_A \},\label{eq:slater}\\
 \varphi_i&=& \phi_{{\bvec{X}}_i}\chi_i\tau_i,\\
 \phi_{{\bvec{X}}_i}({\bvec{r}}_j) & = &  \left(\frac{2\nu}{\pi}\right)^{3/4}
\exp\bigl[-\nu({\bvec{r}}_j-\bvec{X}_i)^2\bigr],
\label{eq:spatial}\\
 \chi_i &=& (\frac{1}{2}+\xi_i)\chi_{\uparrow}
 + (\frac{1}{2}-\xi_i)\chi_{\downarrow}.
\end{eqnarray}
Here ${\cal{A}}$ is the antisymmetrizer, and  $\varphi_i$ is
the $i$th single-particle wave function written by a product of
spatial ($\phi_{{\bvec{X}}_i}$), spin ($\chi_i$), and isospin ($\tau_i$
fixed to be proton or neutron)
wave functions. The width parameter $\nu$ is chosen 
to be  $\nu=0.19$ fm$^{-2}$ for all nucleons as the same as that used for 
AMD+VAP calculations of $^{12}$C and $^{16}$O
in Refs.~\cite{KanadaEn'yo:1998rf,Kanada-Enyo:2017ers}.
Parameters 
${\bvec{Z}}\equiv
\{{\bvec{X}}_1,\ldots, {\bvec{X}}_A,\xi_1,\ldots,\xi_A \}$, which represent  
Gaussian centroid positions and nucleon-spin orientations, are
optimized by the energy variation for each $J^\pi$ state of $\Ne$
so as to minimize the energy expectation value 
$E=\langle \Psi|{\hat H}|\Psi\rangle /\langle \Psi|\Psi\rangle$ with respect to the 
parity and total angular momentum projected wave functions
$\Psi=P^{J\pi}_{MK}\Phi_{\rm AMD}({\bvec{Z}})$.
Here $P^{J\pi}_{MK}$ is the parity and 
total angular momentum projection operator.  

The VAP  is performed for $J^\pi=\{0^+,2^+,4^+\}$, $\{2^-, 3^-,4^-\}$, and $\{1^-, 3^-\}$
by choosing $K=0$, $K=2$, and $K=0$, respectively. For each band, firstly
the band-head state is obtained by the VAP from a randomly chosen initial state, and then higher 
angular momentum states are calculated by the VAP from 
the initial wave function projected from $\Phi_{\rm AMD}({\bvec{Z}})$ obtained for the band-head state. 
Totally eight AMD wave functions $\Phi_{\rm AMD}(\bvec{Z}^{(m)})$ $(m=1,\ldots,8)$ are 
obtained after the VAP, and all of them are 
superposed to obtain final wave functions of the ground and excited states of $\Ne$.
Namely, for the basis wave functions $P^{J\pi}_{MK}\Phi_{\rm AMD}({\bvec{Z}}^{(m)})$ 
projected from the intrinsic wave functions,
the diagonalization of Hamiltonian and norm matrices is done
to obtain $J^\pi$ states. The diagonalization is done
with respect to $K$ and $m$ meaning the $K$-mixing and the configuration $(m)$ mixing. 
As a result of the diagonalization,
$J^\pi=\{0^+,2^+,4^+\}$ states 
in the $K^\pi=0^+_\textrm{hn}$ band are also obtained. 

The effective nuclear interactions used in the present AMD calculation
are the same as those in Refs.~\cite{KanadaEn'yo:1998rf,Kanada-Enyo:2017ers}. 
The MV1 (case 1) central force \cite{TOHSAKI} with the parameters $(b,h,m)=(0,0,0.62)$
and the spin-orbit term of the G3RS force \cite{LS1,LS2} with the strength parameters
$u_{ls}\equiv u_{I}=-u_{II}=3000$ MeV are used.
The Coulomb force is also included.

The AMD calculation of $\Ne$ with this set of interactions obtains  reasonable results of 
energy levels and in-band transitions of the 
$K^\pi=0^+_1$, $K^\pi=0^-_1$, and $K^\pi=0^+_\textrm{hn}$ bands but it gives a higher 
energy of the $K^\pi=2^-$ band than the $K^\pi=0^-_1$ band, which is 
inverse ordering of the $K^\pi=2^-$ and $K^\pi=0^-_1$ bands 
compared with experimental levels.
The excitation energy of the $K^\pi=2^-$ band is sensitive to the strength of spin-orbit interactions
as discussed in Refs.~\cite{Kimura:2003uf,Ohta:2004wm}. 
We can improve the $K^\pi=2^-$ energy with a slight modification of the  
spin-orbit strength
and
obtain the correct ordering of the $K^\pi=2^-$ and $K^\pi=0^-_1$ bands.
In order to discuss possible state mixing between the $K^\pi=2^-$ and $K^\pi=0^-_1$ bands, 
we also use a strength of $u_{ls}=3400$~MeV modified from the original value $u_{ls}=3000$~MeV, 
and perform diagonalization of the basis AMD wave functions already obtained by VAP 
with the default strength ($u_{ls}=3000$~MeV).
We label the AMD calculation with the default and modified strengths, $u_{ls}=3000$~MeV and 
3400 MeV, as AMD and AMD-ls34, respectively. 

In addition to the AMD calculation, a structure calculation of the $\Oa$-cluster model (CM) is 
also performed with GCM. In the GCM framework,  
the Brink-Bloch $\Oa$-cluster wave functions \cite{brink66}
with inter-cluster distances of $1,2\ldots, 10$ fm are superposed. 
In the CM calculation, 
we adopt the same parametrization as that used in the 
$\Oa$-cluster model calculation with the resonating group method in Refs.~\cite{Matsuse73,Matsuse75}.
That is, the width parameter of $\nu=0.16$ fm$^{-2}$ of $^{16}$O and $\alpha$ clusters
and the Volkov No.2 central nuclear interaction with $m=0.62$ are used. 

\subsection{MCC calculation of proton and $\alpha$ scattering off  $\Ne$}

Elastic and inelastic cross sections of  proton and $\alpha$ scattering off $\Ne$ are calculated with
the MCC approach as done in our previous studies of 
Refs.~\cite{Kanada-Enyo:2019prr,Kanada-Enyo:2019qbp,Kanada-Enyo:2019uvg,Kanada-Enyo:2020zpl}.
For the details of the reaction calculations, the reader is referred to those references. 

The nucleon-nucleus potentials are constructed 
in a MFM, where the diagonal and coupling potentials are calculated
by folding the Melbourne $g$-matrix $NN$ interaction \cite{Amos:2000} with 
matter and transition densities of the target nucleus. 
The $\alpha$-nucleus potentials are obtained in an extended nucleon-nucleus
folding (NAF) model \cite{Egashira:2014zda} by folding the calculated
nucleon-nucleus potentials with an $\alpha$ density.

The Melbourne $g$ matrix is an effective $NN$ interaction derived with a bare $NN$ interaction of 
the Bonn B potential~\cite{Mac87}. It contains energy and density dependences
with no adjustable parameter, and works well in application for systematic description of 
proton elastic and inelastic scattering off various nuclei
at energies $E_p=$40--300~MeV ~\cite{Amos:2000,Min10,Toy13,Minomo:2017hjl,Kanada-Enyo:2019uvg,Kanada-Enyo:2020zpl}
and also $\alpha$ elastic and inelastic scattering at energies $E_\alpha=$100--400~MeV
\cite{Egashira:2014zda,Minomo:2016hgc,Kanada-Enyo:2019prr,Kanada-Enyo:2019qbp,Kanada-Enyo:2020zpl}.
In the present calculation of the proton-nucleus potentials, 
the spin-orbit term of the potential is not taken into account to avoid complexity
as in Refs.~\cite{Kanada-Enyo:2019uvg,Kanada-Enyo:2020zpl}. 

As  structure inputs for the target nucleus, the matter $(\rho(r))$ and transition 
$(\rho^\textrm{tr}(r))$ densities of $\Ne$ 
obtained by the AMD and CM calculations are used. 
$J^\pi=0^+$, $1^-$, $2^+$, $3^-$, and $4^+$ 
states in the $K^\pi=0^+_1$,  $K^\pi=2^-$, $K^\pi=0^-_1$,
and $K^\pi=0^+_\textrm{hn}$ bands and $\lambda\le 4$ transitions between them
are adopted in the MCC+AMD calculation, and 
those in the $K^\pi=0^+_1$, $K^\pi=0^-_1$, and $K^\pi=0^+_\textrm{hn}$ bands 
are used in the MCC+CM calculation.
In order to reduce model ambiguity from the structure calculation, 
the theoretical transition densities obtained by the structure calculations
are renormalized in application to the MCC calculations 
so as to fit the experimental transition strengths 
as $\rho^\textrm{tr}(r)\to f^\textrm{tr}\rho^\textrm{tr}(r)$. Here 
the factor  $f^\textrm{tr}$ is determined with the squared ratio 
of experimental ($B_\textrm{exp}(E\lambda)$) 
to theoretical ($B_\textrm{th}(E\lambda)$) strengths as  
$f^\textrm{tr}=\sqrt{B_\textrm{exp}(E\lambda)/B_\textrm{th}(E\lambda)}$ for known values of 
$B_\textrm{exp}(E\lambda)$, and $f^\textrm{tr}=1$ (no renormalization)  is used for unknown cases. 

\section{Structure of  $^{20}$Ne} \label{sec:results1}

\begin{figure}[!h]
\includegraphics[width=6.5 cm]{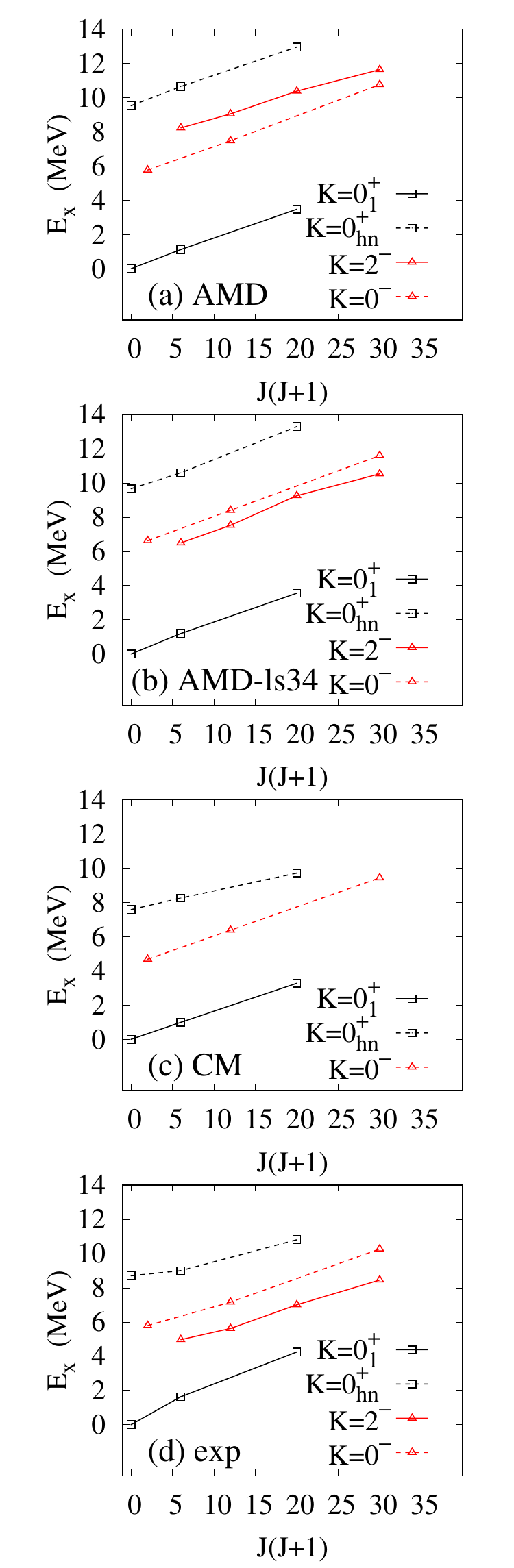}
  \caption{The calculated energy levels of $\Ne$ 
obtained by (a) AMD, (b) AMD-ls34, and (c) CM, 
and (d) the experimental levels assigned to the 
$K^\pi=0^+_1$, $K^\pi=2^-$, $K^\pi=0^-_1$, 
and $K^\pi=0^+_\textrm{hn}$ bands.
  \label{fig:spe}}
\end{figure}

The calculated energy levels of $\Ne$ 
obtained by AMD, AMD-ls34, and CM
are shown in Figs.~\ref{fig:spe}(a), (b), and (c), respectively, 
and the experimental levels assigned to the 
$K^\pi=0^+_1$, $K^\pi=2^-$, $K^\pi=0^-_1$, 
and $K^\pi=0^+_\textrm{hn}$ bands
are shown in Fig.~\ref{fig:spe} (d).
The AMD calculation  reasonably describes the experimental 
energy levels of the $K^\pi=0^+_1$, $K^\pi=0^-_1$, and 
$K^\pi=0^+_\textrm{hn}$ bands (Fig.~\ref{fig:spe}(a)).
However, it overestimates the  $K^\pi=2^-$ levels and gives inverse ordering of the 
$K^\pi=2^-$ and $K^\pi=0^-_1$ bands compared with the experimental levels.
In the AMD-ls34 result with a modified spin-orbit strength (Fig.~\ref{fig:spe} (b)), 
the excitation energy of the $K^\pi=2^-$ band comes down lower than the  $K^\pi=0^-_1$ band, 
and correct ordering of the two negative-parity bands is obtained.
As a result, the state mixing between the $K^\pi=2^-$ and $K^\pi=0^-_1$ bands occurs in the 
$3^-$ states in the AMD-ls34 case. This state mixing is not obtained in the default AMD calculation. 
In the result of CM, the energy levels of the 
$K^\pi=0^+_1$, $K^\pi=0^-_1$, and $K^\pi=0^+_\textrm{hn}$ bands
are reproduced, but the $K^\pi=2^-$ band is missing 
because the $\Oa$ model space contains only axial symmetric configurations
with pure $K=0$ components. 

\begin{table}[ht]
\caption{Excitation energies ($E_x$) and root-mean-square matter radii  ($R$) of $^{20}$Ne.
The calculated values obtained with the AMD and CM calculations 
and the experimental values are listed.
The experimental value of the point-proton rms radius of the ground state is $R=2.888(2)$~fm 
from the charge radius data\cite{Angeli2013}.
 \label{tab:radii}
}
\begin{center}
\begin{tabular}{lrrrrrrrrccccc}
\hline
\hline
\multicolumn{2}{c}{exp}     &    \multicolumn{2}{c}{AMD} &    \multicolumn{2}{c}{CM} \\
  $J^\pi$  & $E_x$~(MeV)&   $E_x$~(MeV)  & $R$~(fm) &   $E_x$~(MeV) & $R$~(fm) \\
$K^\pi=0^+_1$\\
$0^+_1$	&	0	&	0.0 	&	3.01 	&	0.0 	&	2.95 	\\
$2^+_1$	&	1.634	&	1.1 	&	3.01 	&	1.0 	&	2.94 	\\
$4^+_1$	&	4.248	&	3.5 	&	2.99 	&	3.3 	&	2.92 	\\
$K^\pi=2^-$\\		
$2^-_1$	&	4.967	&	8.2 	&	2.94 	&		&		\\
$3^-_1$	&	5.621	&	9.0 	&	2.96 	&		&		\\
$4^-_1$	&	7.004	&	10.4 	&	2.97 	&		&		\\
$K^\pi=0^-_1$\\		
$1^-_1$	&	5.788	&	5.8 	&	3.20 	&	4.7 	&	3.20 	\\
$3^-_2$	&	7.156	&	7.5 	&	3.19 	&	6.4 	&	3.22 	\\
\hline
\hline
\end{tabular}
\end{center}
\end{table}

\begin{table*}[ht]
\caption{$E2$ transition strengths for 
in-band transitions in the $K^\pi=0^+_1$, $K^\pi=2^-$, and $K^\pi=0^-_1$ bands, 
$E\lambda$ and isoscalar dipole (IS1) transition strengths to the ground state, and electric quadrupole moment ($Q$) of the
$2^+_1$ state.
The theoretical values 
obtained by the AMD~(default), AMD-ls34~(modified spin-orbit strength), and CM calculations
are listed together with the experimental values 
from Refs.~\cite{Hausser:1971mml,Tilley:1998wli}. 
Theoretical values of the 
($\Oa$)+($\CBe$) coupled-channel OCM (CC-OCM) \cite{Fujiwara79c} and the 
deformed-basis AMD (def-AMD) \cite{Kimura:2003uf} are also shown.
In addition, the $B(E\lambda)$ values reduced from 
inelastic scattering data of $(e,e')$~\cite{Horikawa:1971oau}, 
$(p,p')$ at 800~MeV\cite{Blanpied:1988ew}, 24.5 MeV~\cite{deSwiniarski:1976rcl,Swiniarski72}, and $(d,d')$ 
at 52 MeV~\cite{Hinterberger:1968odq} are also shown.
The units are $e^2$fm$^{2\lambda}$ for $B(E\lambda)$, $e^2$fm$^{6}$ for $B(\textrm{IS1})$, and $e$fm$^{2}$ for $Q$.
 \label{tab:BEl}
}
\begin{center}
\begin{tabular}{lrrrrrrrrrrrrrccccc}
\hline
\hline
	&	Exp.\cite{Tilley:1998wli,Hausser:1971mml} &	CM	&	AMD	&	AMD-ls34	&		CC-OCM\cite{Fujiwara79c}	&	def-AMD\cite{Kimura:2003uf} 	\\
$B(E2;2^+_1\to 0^+_1)$	&	65.4(3.2)	&	53 	&	69 	&	63 	&	57.0	&	70.3	\\
$B(E2;4^+_1\to 2^+_1)$	&	71(6)	&	67 	&	92 	&	84 	&	70.9	&	83.7	\\
\ \\													
$B(E2;3^-_1\to 2^-_1)$	&	113(29)	&		&	95 	&	89 	&	107.5	&	102.8	\\
$B(E2;4^-_1\to 3^-_1)$	&	77(16)	&		&	79 	&	68 	&	77.0	&	77.8	\\
$B(E2;4^-_1\to 2^-_1)$	&	34(6)	&		&	32 	&	31 	&	34.0	&	38.5	\\
\ \\													
$B(E2;3^-_2\to 1^-_1)$	&	161(26)	&	178 	&	163 	&	150 	&		&	151.2	\\
\ \\													
$B(E3;3^-_1\to 0^+_1)$	&	260(90)	&		&	53 	&	155 	&	29.9	&		\\
$B(E3;3^-_2\to 0^+_1)$	&		&	543 	&	548 	&	335 	&		&		\\
$B(IS1;1^-_1\to 0^+_1)/4$	&		&	222 	&	164 	&	129 	&		&		\\
$B(E4;4^+_1\to 0^+_1)$	&		&	5060 	&	9270 	&	7440 	&		&		\\
\ \\													
$Q(2^+_1)$	&	$-23(3)$	&	$-14.7$	&	$-16.9$	&	$-16.0$	&	$-15.2$	&		\\
\hline																							
	&	$(e,e')$\cite{Horikawa:1971oau}	&	$(p,p')$\cite{Blanpied:1988ew}	&	$(p,p')$\cite{deSwiniarski:1976rcl,Swiniarski72}	&	$(d,d')$\cite{Hinterberger:1968odq}	&		&			&		\\
	&		&	800 MeV	&	24.5 MeV	&	52 MeV	&		&		&			\\
$B(E2;2^+_1\to 0^+_1)$	&	71 	&	52	&	66	&		&		&		&			\\
$B(E4;4^+_1\to 0^+_1)$	&	8100 	&	5530	&	15200	&		&		&			&		\\
$B(E3;3^-_1\to 0^+_1)$	&		&	300	&	450	&	420 	&		&		&			\\
$B(E3;3^-_2\to 0^+_1)$	&		&	146	&	230	&	450 	&		&		&			\\
\hline
\hline
\end{tabular}
\end{center}
\end{table*}

\begin{figure}[!h]
\includegraphics[width=9.5 cm]{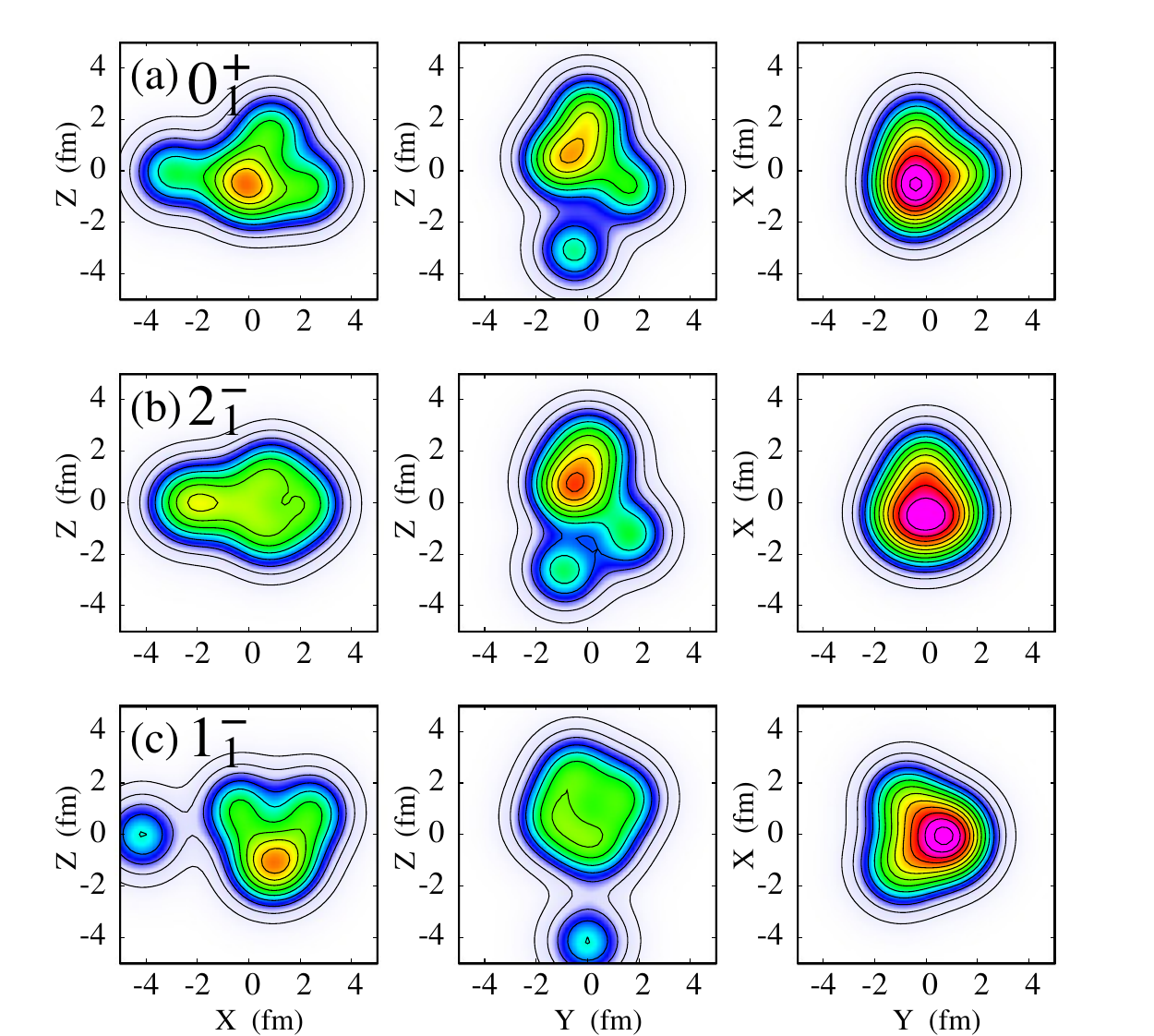}
  \caption{
Density distribution of intrinsic wave functions 
for the band-head states, (a) $0^+_1$, (b) $2^-_1$, and (c) $1^-_1$ of the 
$K^\pi=0^+_1$, $K^\pi=2^-$, and $K^\pi=0^-_1$ bands
of $\Ne$ obtained with AMD.
The density projected onto $X$-$Z$, 
$Y$-$Z$, and $Y$-$X$ planes are shown in left, middle, and right panels, respectively. 
Intrinsic axises are chosen as 
$\langle ZZ\rangle\ge \langle YY\rangle\ge \langle XX\rangle$ and 
 $\langle XY\rangle=\langle YZ\rangle=\langle ZX\rangle=0$.
  \label{fig:dense-cont}}
\end{figure}

In Table \ref{tab:radii}, the calculated 
values of excitation energies ($E_x$) and root-mean-square radii ($R$) of $\Ne$
obtained with AMD and CM are listed together with the experimental excitation energies. 
Experimentally, the $J^\pi=\{0^+_1, 2^+_1, 4^+_1\}$, $\{2^-_1, 3^-_1, 4^-_1\}$, and $\{1^-_1, 3^-_2\}$ 
states with strong in-band $E2$ transitions
are assigned to 
the $K^\pi=0^+_1$,  $K^\pi=2^-$, and $K^\pi=0^-_1$ bands, respectively.
Following this experimental assignment, we use the label $3^-_1$ ($3^-_2$) for the 
$K^\pi=2^-$($K^\pi=0^-_1$) band member of the theoretical results.
The negative-parity states in the $K^\pi=0^-_1$ band have large radii 
compared to the
$K^\pi=0^+_1$ and $K^\pi=2^-$ bands because of a spatially developed $\Oa$ structure.

Figure ~\ref{fig:dense-cont} shows 
intrinsic density distribution of the basis AMD wave functions 
for the band-head  states, 
$0^+_1$($K^\pi=0^+_1$), $2^-_1$($K^\pi=2^-$), and $1^-_{1}$($K^\pi=0^-_1$).
In the three states, one or two $\alpha$ clusters are formed. The $0^+_1$($K^\pi=0^+_1$) state shows an $\Oa$ like structure,
whereas the $1^-_{1}$($K^\pi=0^-_1$) state has the most prominent $\Oa$ cluster structure.
Qualitatively, 
these two bands can be regarded as the  parity doublets constructed 
from the $\Oa$-cluster structure as in a simple $\Oa$-cluster model, but strictly
speaking it is not correct because 
the $0^+_1$($K^\pi=0^+_1$) state contains a deformed $^{16}$O cluster showing a significant component of 
internal excitation of the cluster. 
The $K^\pi=2^-$ band shows a $\Caa$-like structure with an axial asymmetric shape, where
two $\alpha$ clusters are formed around a $^{12}$C cluster. 
As a result of formation of the $^{12}$C cluster,  the $K^\pi=2^-$ band gains 
the spin-orbit interaction.
Compared to the $K^\pi=0^-_1$ band, the $K^\pi=2^-$ band has a compact structure
with a mean-field aspect of particle-hole excitation on the prolate state.
It means that the $K^\pi=2^-$ band has the duality of cluster and mean-field features.

The results of electric ($E\lambda)$ and isoscalar dipole (IS1) transition strengths 
and electric quadrupole moment ($Q$) are shown in Table \ref{tab:BEl}.
The theoretical values obtained by AMD, AMD-ls34, and CM 
are shown together with the experimental data 
from Refs.~\cite{Tilley:1998wli,Hausser:1971mml}. Moreover, 
theoretical values of the CC-OCM~\cite{Fujiwara79c} 
and def-AMD~\cite{Kimura:2003uf} calculations are also shown for comparison. 
In addition to the experimental $B(E\lambda)$ measured by $\gamma$ decays, 
the values reduced from $(e,e')$ scattering data~\cite{Horikawa:1971oau} 
and those evaluated from inelastic scattering of 
$(p,p')$ \cite{Blanpied:1988ew,deSwiniarski:1976rcl,Swiniarski72} and $(d,d')$ 
\cite{Hinterberger:1968odq} are also shown in Table \ref{tab:BEl}. 
Note that uncertainty remains in the evaluation with hadron scattering because it
relies on the phenomenological reaction analysis 
and shows significant dependences on energy and projectile. 

The observed in-band $E2$ transitions in 
the $K^\pi=0^+_1$ $K^\pi=2^-$, and $K^\pi=0^-_1$ bands
are reproduced well by the ADM calculation. The agreement is almost
the same quality as other theoretical calculations of CC-OCM and def-AMD.
The experimental $Q$ moment of the $2^+_1$ state is somewhat underestimated by the AMD and CM 
calculations. 

For the $E4$ transition in the $K^\pi=0^+_1$ band, the strength $B(E4;4^+_1\to 0^+_1)$ obtained with 
the AMD calculation is consistent with the $(e,e')$ scattering, 
while the CM calculation gives a weaker $E4$ transition.
The values evaluated from $(p,p')$ scattering strongly depend on energies and
have large uncertainty.
For $E3$ transitions from the $K^\pi=0^-_1$ band, AMD and
CM give the remarkably strong $3^-_2\to 0^+_1$ transition
because of the developed $\Oa$-cluster structure.
For the transition from the $K^\pi=2^-$ band, the AMD calculation obtains 
the weak $3^-_1\to 0^+_1$ transition 
as one order of magnitude smaller strength as the 
$3^-_2\to 0^+_1$ strength.
These AMD results of $B(E3;3^-_1\to 0^+_1)$ and $B(E3;3^-_2\to 0^+_1)$
are consistent with the CC-OCM calculation, but not consistent with the observation.
The observed $B(E3;3^-_1\to 0^+_1)$ is much larger than the AMD and CC-OCM results. 
Moreover, the evaluation from $(p,p')$ and $(d,d')$ scattering
suggests the same order transitions to the $3^-_1$($K^\pi=2^-$) and $3^-_2$($K^\pi=0^-_1$) states
though uncertainty still remains.

Transition properties from the $0^+_1$ state 
to the $3^-_1$($K^\pi=2^-$) and $3^-_2$($K^\pi=0^-_1$) states
are sensitive to the state mixing between the $K^\pi=2^-$ and $K^\pi=0^-_1$
bands. Let us discuss effects of the state mixing on the $E3$ transition strengths 
by comparing the AMD-ls34 and AMD results for the cases with and without state mixing, 
respectively. 
As shown in Table \ref{tab:BEl}, 
the $B(E3;3^-_1\to 0^+_1)$ value of AMD-ls34 is as three times large as that of AMD
because of mixing of the $K^\pi=0^-_1$ cluster component in the $3^-_1$ state. 
On the other hand, 
the $B(E3;3^-_2\to 0^+_1)$ value decreases in the AMD-ls34 result compared to AMD
because of destructive mixing of the 
$K^\pi=2^-$ component in the $3^-_2$ state.
In order to describe the experimental
$B(E3;3^-_1\to 0^+_1)$, 
the state mixing case of AMD-ls34 seems more likely than the almost no mixing case of AMD.
Such the significant mixing may originate 
in coupling of the $\Caa$ and $\Oa$ cluster structures contained in the $K^\pi=2^-$ and $K^\pi=0^-_1$
bands, respectively, as follows. The $\Oa$ cluster structure can be smoothly 
transformed into the $\Caa$ with internal excitation of the $^{16}\textrm{O}$ cluster. Owing to this
cluster degree of freedoms between the $\Oa$ and $\Caa$ channels, 
the $3^-$ excitations in two channels can couple with each other.

\begin{figure}[!h]
\includegraphics[width=6 cm]{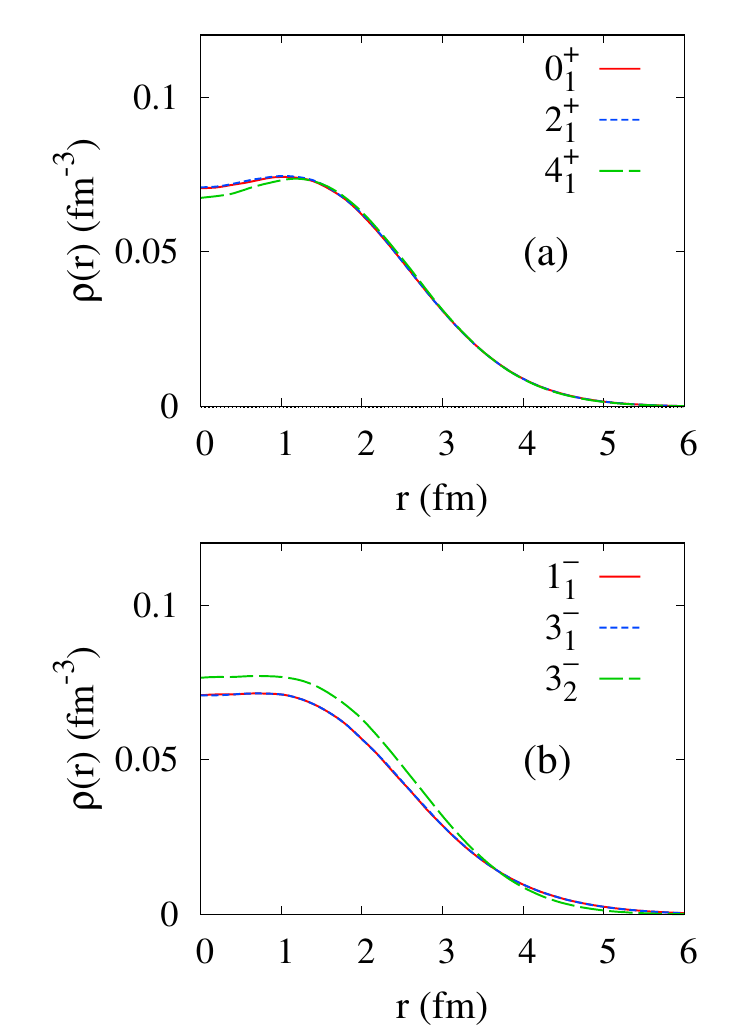}
  \caption{Matter densities of $\Ne$ calculated with AMD.
  \label{fig:dense}}
\end{figure}

\begin{figure}[!h]
\includegraphics[width=9 cm]{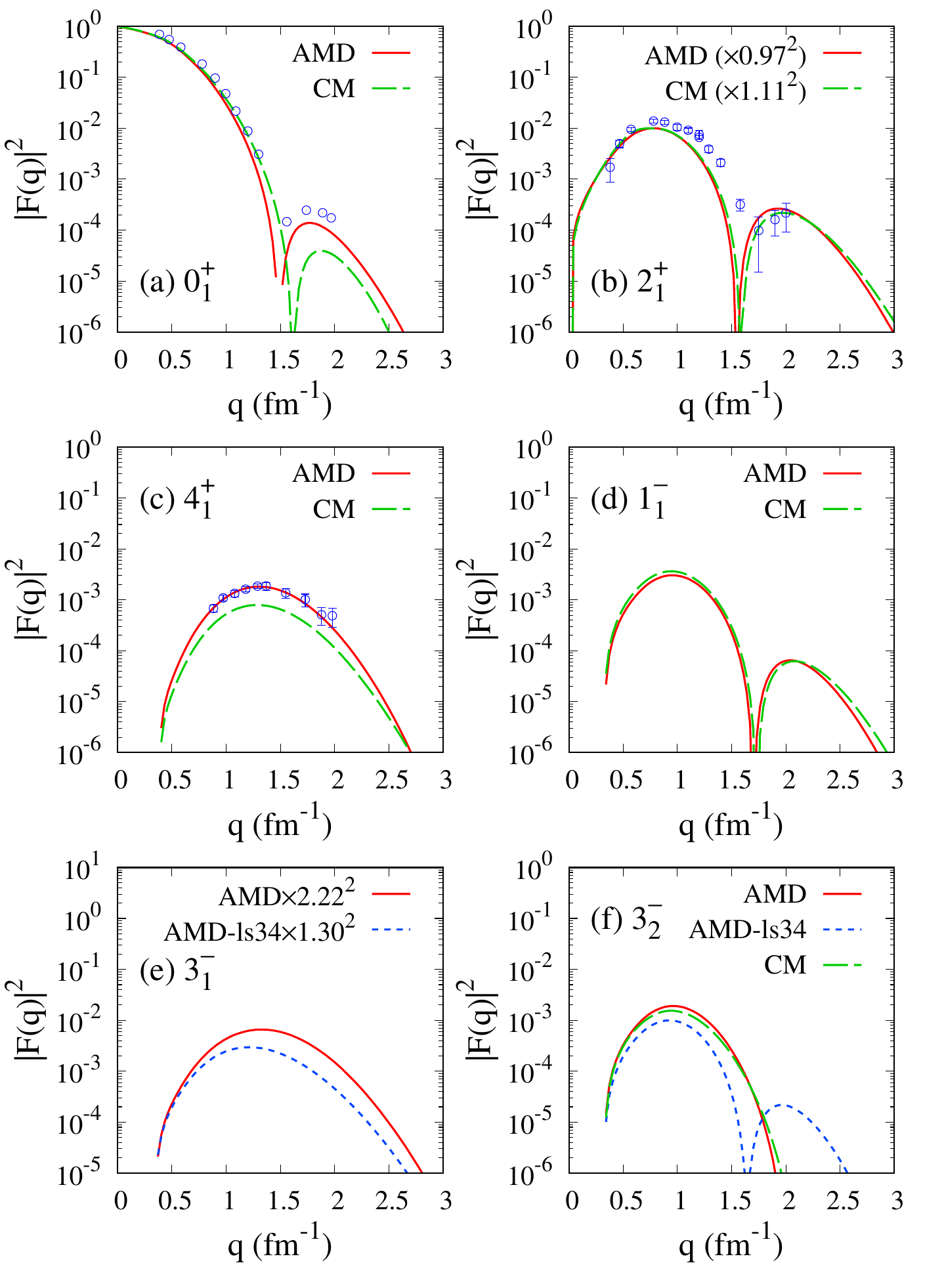}
  \caption{Elastic and inelastic form factors of $\Ne$ calculated with 
AMD (red solid lines) and CM (green dashed lines) 
compared with the experimental data (circles).
The $3^-_1$ and $3^-_2$
form factors calculated with AMD-ls34 (blue dotted lines) 
are also shown in panels (e) and (f), respectively. 
The theoretical $2^+_1$ form factors obtained by AMD (CM) are 
renormalized by $f^\textrm{tr}=0.97$ (1.11).   
The theoretical $3^-_1$ form factors obtained by AMD and AMD-ls34 are 
multiplied by $f^\textrm{tr}=2.22$ and 1.30, respectively. 
The experimental data are those measured by electron scattering \cite{Horikawa:1971oau}.
  \label{fig:form}}
\end{figure}

\begin{figure}[!h]
\includegraphics[width=6 cm]{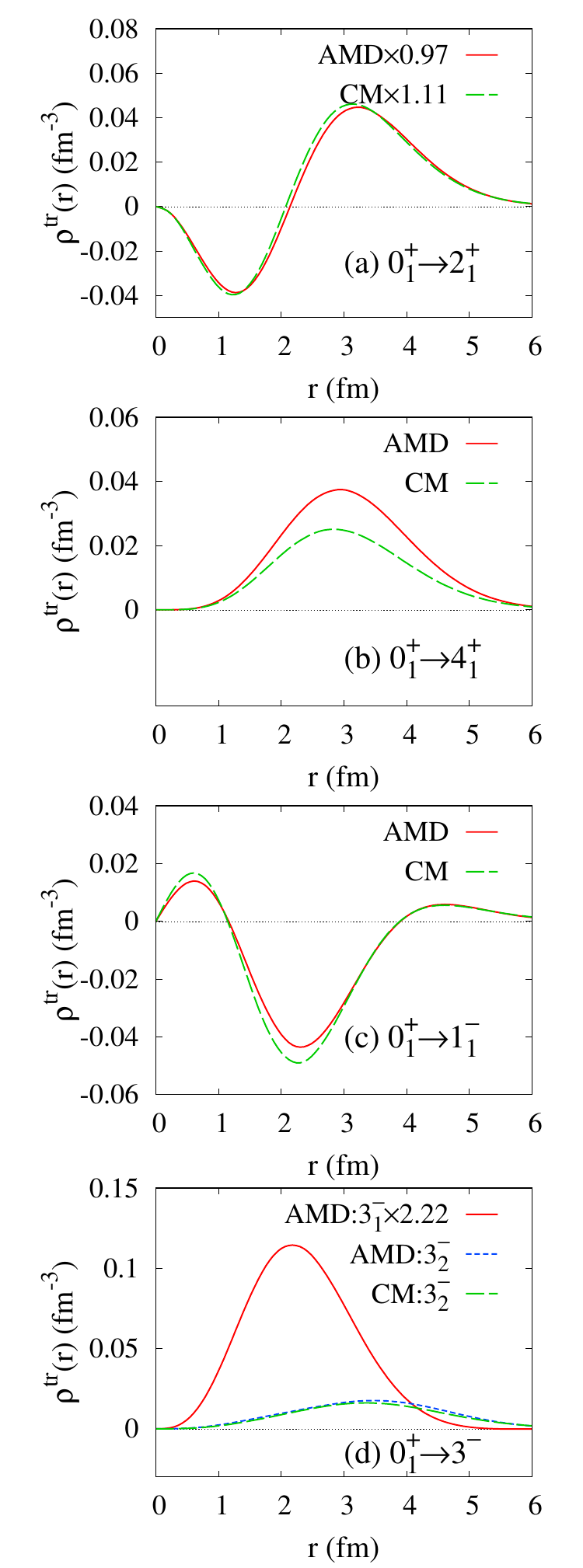}
  \caption{
Transition densities of $\Ne$ calculated with AMD and CM.
Theoretical values obtained by AMD (CM) for the $2^+_1$ state are
renormalized by $f^\textrm{tr}=0.97$ (1.11), and those for  $3^-_1$ state are 
multiplied by $f^\textrm{tr}=2.22$. 
  \label{fig:trans}}
\end{figure}

The calculated matter densities are shown in Fig.~\ref{fig:dense}. 
Compared to the  $K^\pi=0^+_1$ and  $K^\pi=2^-$ bands, 
the $1^-_1$ and $3^-_2$ states in the $K^\pi=0^-$ band show relatively 
broader density distribution 
in the outer region
because of the developed $\Oa$ cluster structure, 
but the state dependence of matter densities is not so strong.
The calculated form factors and transition densities 
are shown in Figs.~\ref{fig:form} and \ref{fig:trans}, respectively. 
For the $2^+_1\to 0^+_1$ and $3^-_1\to 0^+_1$ transitions, 
theoretical values are renormalized 
with $f^\textrm{tr}$ determined from the experimental and theoretical $B(E\lambda)$ values
listed in Table \ref{tab:BEl}.
The experimental form factors observed by $(e,e')$ scattering are also shown
in Figs.~\ref{fig:form}(a), (b), and (c). 
The AMD and CM calculations reproduce the elastic form factors, and also 
describe the inelastic form factors to the $2^+_1$ state.
The calculated $4^+_1$ form factors obtained with 
AMD are in good agreement with the $(e,e')$ data, 
but those with CM 
underestimates the data (see Fig.~\ref{fig:form}(c)). 

In the AMD result for $E3$ transitions, clear differences between the $3^-_1$ and $3^-_2$ states
can be seen in the form factors and transition densities, 
which are shown by red lines of Figs.~\ref{fig:form}(e), \ref{fig:form}(f), and \ref{fig:trans}(d). 
The $3^-_1$ form factors have the higher peak at a larger $q$, while the $3^-_2$ form factors 
show the lower peak at a smaller $q$. 
Similarly, one can see the difference also in the transition densities:
narrower distributions of the $3^-_1\to 0^+_1$ transition densities and 
broader distributions with the outer tail of the $3^-_2\to 0^+_1$ transition densities 
because of the developed $\Oa$ cluster structure.
One can say again that the $E3$ form factors and transition densities
are sensitive to the state mixing between the $K^\pi=2^-$ and $K^\pi=0^-_1$ bands.
Detailed discussions of its effect are given 
in Sec.~\ref{sec:discussions}.


\section{Proton and $\alpha$ scattering}  \label{sec:results2}

The MCC calculations with AMD and CM are performed for proton scattering 
at incident energies of $E_p=25$, $30$, and $35$~MeV and 
$\alpha$ scattering at $E_\alpha=104$, 146, and 386~MeV. 
To see the coupled channel (CC) effect, the one-step calculation of the distorted wave born approximation (DWBA) is 
also performed using the AMD densities.
As described previously, the theoretical transition densities are 
renormalized by multiplying the factors ($f^\textrm{tr}$), which are determined as 
$f^\textrm{tr}=\sqrt{B_\textrm{exp}(E\lambda)/B_\textrm{th}(E\lambda)}$ to 
fit the experimental data of $B(E2;2^+_1\to 0^+_1)$, $B(E2;4^+_1\to 2^+_1)$, $B(E2;3^-_2\to 1^-_1)$, and 
$B(E3;3^-_1\to 0^+_1)$. For the $E2;2^+_1\to 2^+_1$ transition, $f^\textrm{tr}$ is chosen to adjust the theoretical $Q(2^+_1)$ to the 
experimental value. For other transitions, $f^\textrm{tr}=1$~(no renormalization) is used. 

\begin{figure*}[!h]
\includegraphics[width=18. cm]{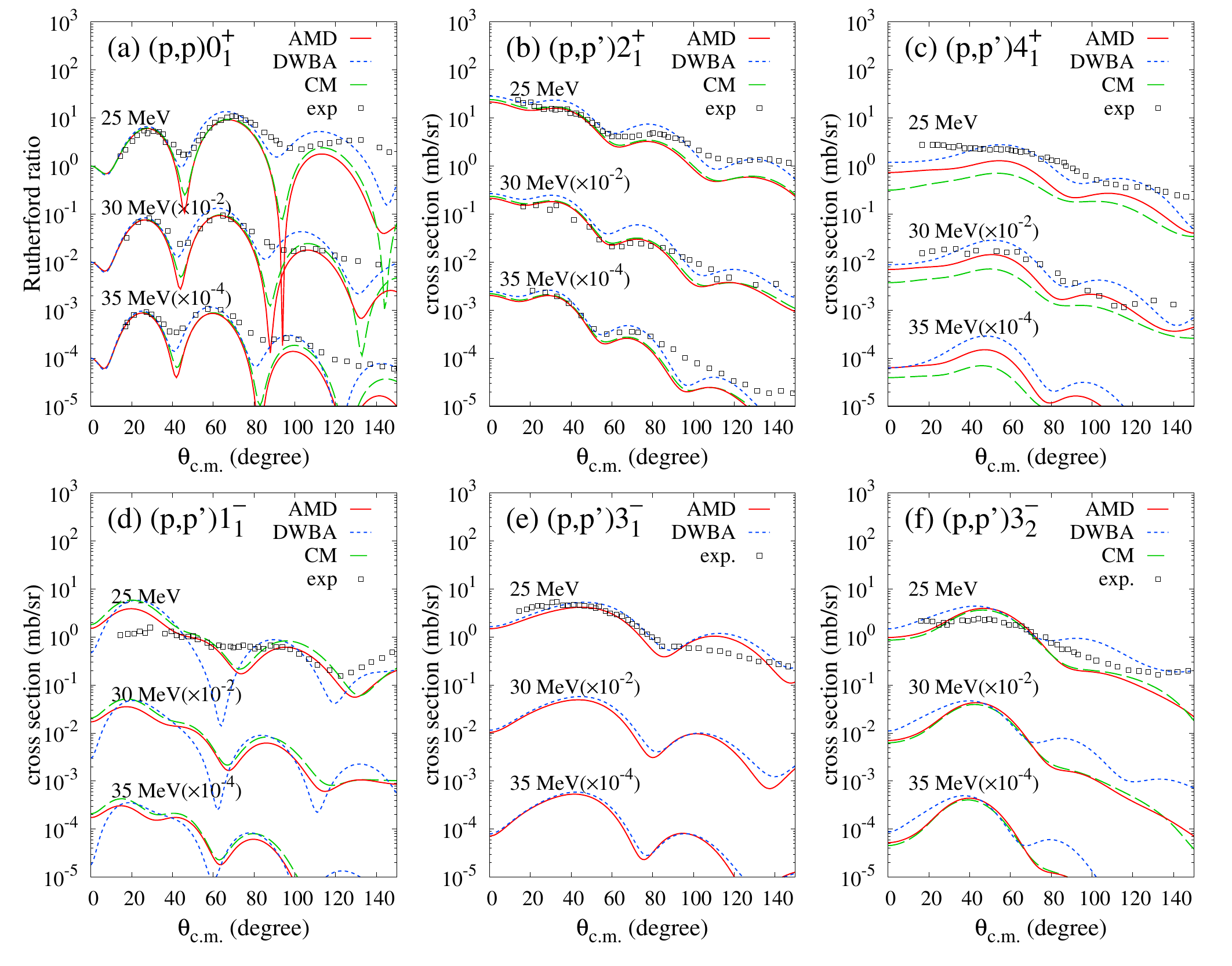}
  \caption{
Cross sections of proton elastic and inelastic scattering off $\Ne$ 
at incident energies of $E_p=25$, 30, and 35 MeV calculated with 
MCC+AMD (red solid lines),  DWBA+AMD (blue dotted lines), and MCC+CM (green dashed lines), 
which are labeled as 
AMD, DWBA, and CM, respectively.
Experiment data are cross sections at 
$E_p=24.5$~MeV~\cite{deSwiniarski:1976rcl,Swiniarski72}, 
30~MeV~\cite{deSwiniarski:1976rcl}, and 35~MeV~\cite{Fabrici:1980zz}.
  \label{fig:cross-ne20p}}
\end{figure*}

The calculated cross sections of proton elastic and inelastic scattering are shown  
in Fig.~\ref{fig:cross-ne20p} compared with the experimental data. 
The MCC+AMD (red solid lines) and MCC+CM (green dashed lines) 
reproduce the $0^+_1$ and $2^+_1$ cross sections data well at the first and second peaks.
For the $4^+_1$ cross sections, the observe data do not show clear peak structures enough to discuss 
diffraction patterns. The MCC+AMD reasonably reproduces the global amplitudes of 
the $4^+_1$ data, while the MCC+CM gives smaller $4^+_1$ cross sections 
than MCC+AMD and the data because of the weaker $E4$ transition than AMD result.
The  $1^-_1$  cross sections are reasonably described 
with the MCC+AMD and MCC+CM calculations except for forward angles. 
As for the $3^-_1$ cross sections, the MCC+AMD reproduces the 
first peak amplitude of the data, but somewhat overestimates the second peak amplitude. 
In comparison of the DWBA+AMD (blue dotted lines) and MCC+AMD (red solid lines) 
calculations, non-negligible CC effects are seen in this energy range $E_p=25$--35~MeV 
except for the $3^-_1$ state.

\begin{figure*}[!h]
\includegraphics[width=18. cm]{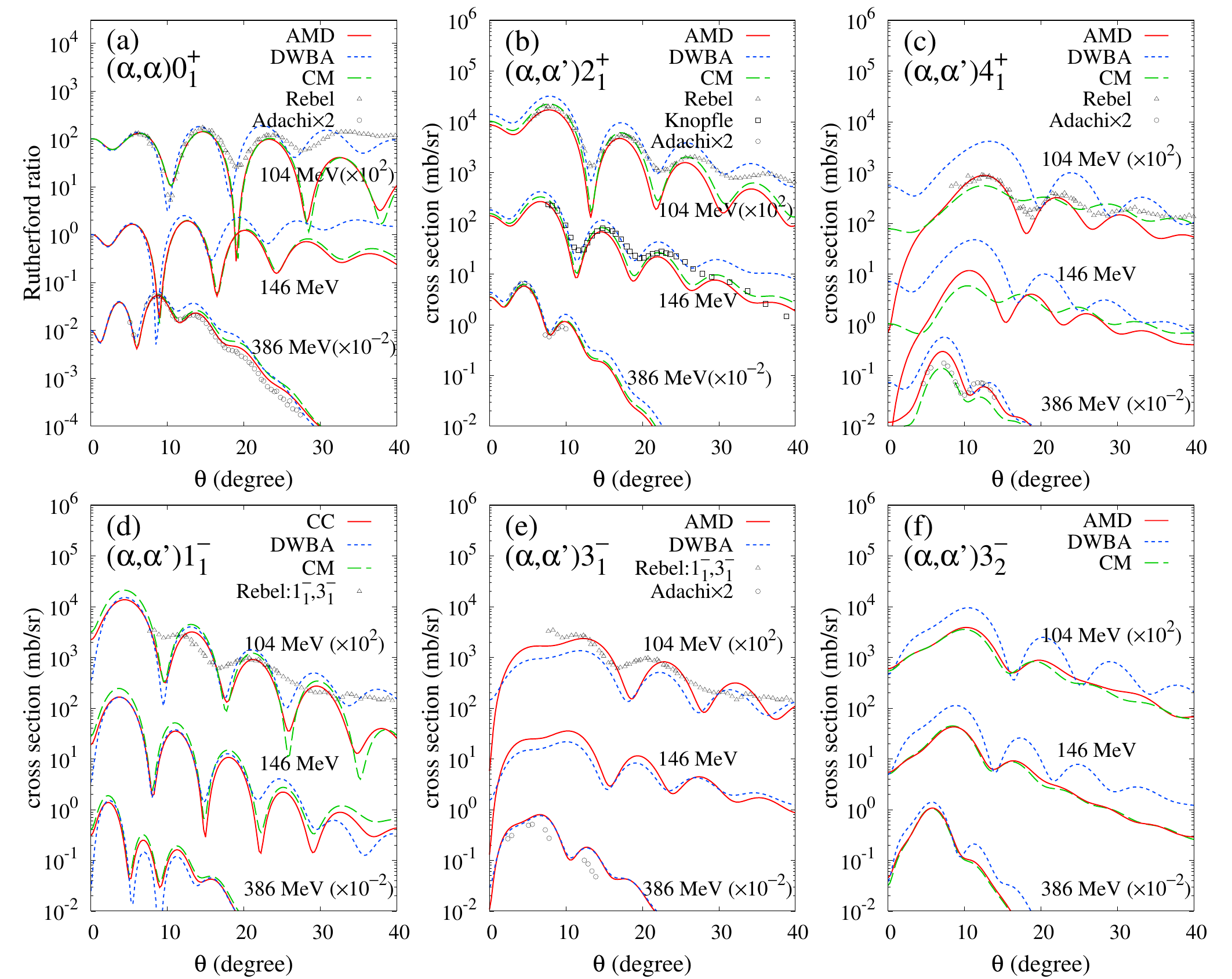}
  \caption{ 
Cross sections of $\alpha$ elastic and inelastic scattering off $\Ne$ at incident energies of $E_\alpha=104$, 146, and 
386 MeV calculated with 
MCC+AMD (red solid lines),  DWBA+AMD (blue dotted lines), and MCC+CM (green dashed lines), 
which are labeled as 
AMD, DWBA, and CM, respectively.
Experiment data are cross sections 
at $E_\alpha=$104~MeV \cite{Rebel:1972nip},  and 146~MeV\cite{Knopfle:1976npt}, and  
386 MeV \cite{Adachi:2018pql}. The $E_\alpha=386$~MeV data 
from Ref.~\cite{Adachi:2018pql} are multiplied by a factor of two.
The $(\alpha,\alpha')$ data 
at $E_\alpha=$104~MeV in the panels (d) and (e) are the 
cross sections observed for $E_x=5.7$~MeV and may contain $1^-_1$(5.79~MeV) and 
$3^-_1$(5.62~MeV) contributions. 
  \label{fig:cross-ne20a}}
\end{figure*}

The $\alpha$ elastic and inelastic cross sections are shown in  Fig.~\ref{fig:cross-ne20a}. 
The calculated cross sections are compared with the experimental data. 
For the elastic scattering (Fig.~\ref{fig:cross-ne20a}(a)), 
the $E_\alpha=104$~MeV data from Ref.~\cite{Rebel:1972nip}
are reproduced well by MCC+AMD (red solid lines) and
MCC+CM (green dashed lines) except for backward angles, 
whereas the $E_\alpha=386$~MeV data from Ref.~\cite{Adachi:2018pql} are about two times smaller than
the present MCC calculations. We do not know the reason for this apparent inconsistency, but 
uncertainty 
from the present reaction model is unlikely because its applicability to 
$\alpha$ elastic scattering at $E_\alpha=$100--400~MeV has been already examined for various target nuclei \cite{Egashira:2014zda,Minomo:2016hgc,Kanada-Enyo:2019prr,Kanada-Enyo:2019qbp,Kanada-Enyo:2019uvg,Kanada-Enyo:2020zpl}.
Therefore, it is likely that 
the $E_\alpha=386$~MeV data in Ref.~\cite{Adachi:2018pql} contains 
uncertainty of the normalization. Assuming the normalization to be 
an overall factor of two, 
we multiply the original $(\alpha,\alpha)$ and $(\alpha,\alpha')$ 
data of Ref.~\cite{Adachi:2018pql} by this factor, 
and obtain excellent agreement of the calculations with the $(\alpha,\alpha)$ data
as shown in Fig.~\ref{fig:cross-ne20a}(a). 
Also the $2^+_1$ cross sections at $E_\alpha=104$--$386$~MeV are 
 well reproduced by the MCC+AMD and MCC+CM calculations. 
For the $4^+_1$ cross sections, the MCC+AMD result seems better than the MCC+CM result. 
For the $1^-_1$ and $3^-_1$ cross sections, there is no data for individual states
at $E_\alpha=104$~MeV, but the cross sections at $E_x=5.7$ MeV were reported by 
the $E_\alpha=104$~MeV experiment in Ref.~\cite{Rebel:1972nip}. The data  may 
contain the $1^-_1$(5.79 MeV) and $3^-_1$(5.62 MeV) contributions, 
but they can not be described by 
a simple sum of the calculated  $1^-_1$ and $3^-_1$ cross sections of the present calculations.
The $E_\alpha=386$ MeV data of the $3^-_1$ cross sections are 
somewhat overestimated by the MCC+AMD calculation, in particular, at the second peak. 
In comparison of the MCC+AMD (red solid lines) and DWBA+AMD (blue dotted lines) 
calculations for $\alpha$ scattering, one can see 
non-negligible CC effect, in particular, at $E_\alpha=104$ MeV, but the CC effect
becomes weaker at $E_\alpha=386$ MeV.

\section{Discussion} \label{sec:discussions}

As discussed in previous sections, the structure calculation of AMD-ls34 with the modified 
spin-orbit strength 
suggests possible state mixing between the $K^\pi=2^-$ and $K^\pi=0^-_1$ bands 
in the $3^-_1$  and $3^-_2$ states of $\Ne$.
Let us remind that the default AMD calculation gives almost no 
state mixing and obtains the theoretical value of  $B(E3;3^-_1\to 0^+_1)$=53~$e^2\textrm{fm}^4$
much smaller than the experimental value of  $260\pm 90$~$e^2\textrm{fm}^4$. 
In the AMD-ls34 result, the $K^\pi=2^-$ band comes down
to the lower energy than the $K^\pi=0^-_1$ band consistently with the experimental energy spectra, 
and the state mixing occurs between the $3^-_1$($K^\pi=2^-$)  and $3^-_2$($K^\pi=0^-_1$) states.
As a consequence of mixing of the $K^\pi=0^-_1$ cluster component in the $3^-_1$($K^\pi=2^-$) state, 
the theoretical $B(E3;3^-_1\to 0^+_1)$ value is enhanced to be 
$B(E3;3^-_1\to 0^+_1)=155$~$e^2\textrm{fm}^4$ being in better agreement with the experimental value. 

\begin{figure}[!h]
\includegraphics[width=6 cm]{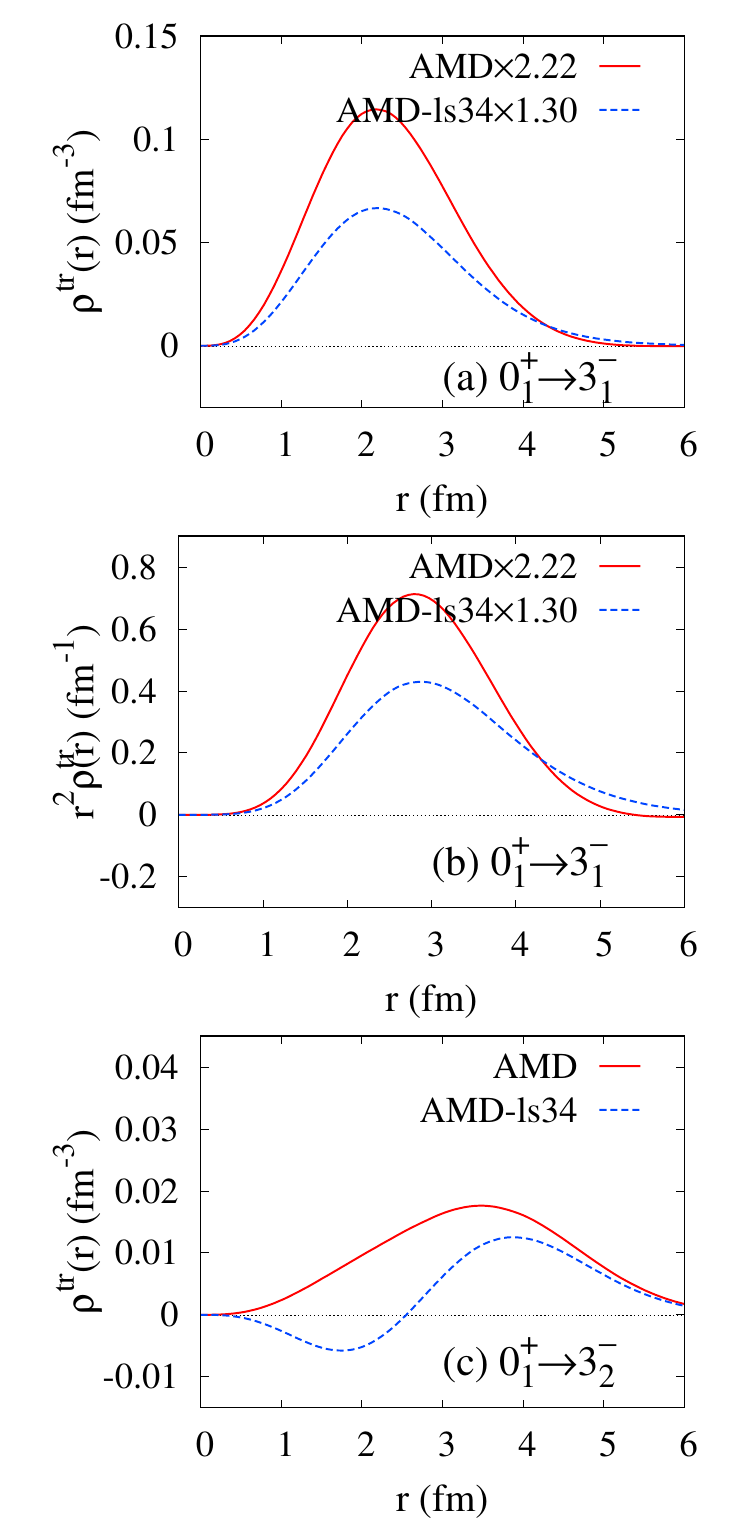}
  \caption{
Transition densities from the ground to $3^-$ states obtained with AMD (default) 
and AMD-ls34 (modified spin-orbit strength). 
(a) transition densities to the $3^-_1$ state, (b) those but $r^2$-weighted, (b)
transition densities to the $3^-_2$ state.
Transition densities to the $3^-_1$ state in (a) and (b) are 
are renormalized by $f_\textrm{tr}=2.22$ for AMD and 1.30 for AMD-ls34.
  \label{fig:trans-ls}}
\end{figure}

This state mixing between the $3^-_1$  and $3^-_2$ states affects the 
$E3$ form factors and transition densities. The AMD (red solid lines) and AMD-ls34 (blue dotted lines)
results for the form factors are compared 
in Figs.~\ref{fig:form}(e) and (f), and those for the transition densities are compared in 
Fig.~\ref{fig:trans-ls}.
In these figures, the $3^-_1$ form factors and transition densities are renormalized 
with $f^\textrm{tr}=2.22$ for AMD and $f^\textrm{tr}=1.30$ for AMD-ls34
so as to fit the data of $B(E3;3^-_1\to 0^+_1)$. 
Namely, the renormalized $3^-_1$ transition densities 
of the two calculations (AMD and AMD-ls34) in Fig.~\ref{fig:trans-ls} 
give the same value of $B(E3;3^-_1\to 0^+_1)=260$~$e^2\textrm{fm}^4$. 
Nevertheless, behaviors of the transition densities are different between the AMD and AMD-ls34 results.
For the $3^-_1$ transition densities (Figs.~\ref{fig:trans-ls}(a) and (b)), AMD-ls34 gives 
a lower peak amplitude in the inner region ($r=$2--3 fm) and a longer tail in the outer region  ($r\sim 5$ fm)
because of the mixing of the $K^\pi=0^-_1$ cluster component. 
As a result, in the $3^-_1$ form factors of AMD-ls34, the peak amplitude gets smaller and 
the peak position shifts to a smaller $q$ (see the blue dotted line of Fig.~\ref{fig:form}(e)). 
Also the $3^-_2$ transition densities are strongly affected by the state mixing
as shown in Fig.~\ref{fig:trans-ls}(c).
Because of the destructive mixing of the $K^\pi=2^-$ component, 
inner amplitudes in the $r<3$ fm region are suppressed and the outer peak around $r=3$--4 fm 
gets smaller and shifts outwards in AMD-ls34~(blue dotted lines) 
than in AMD~(red solide lines). 
 
\begin{figure*}[!h]
\includegraphics[width=13 cm]{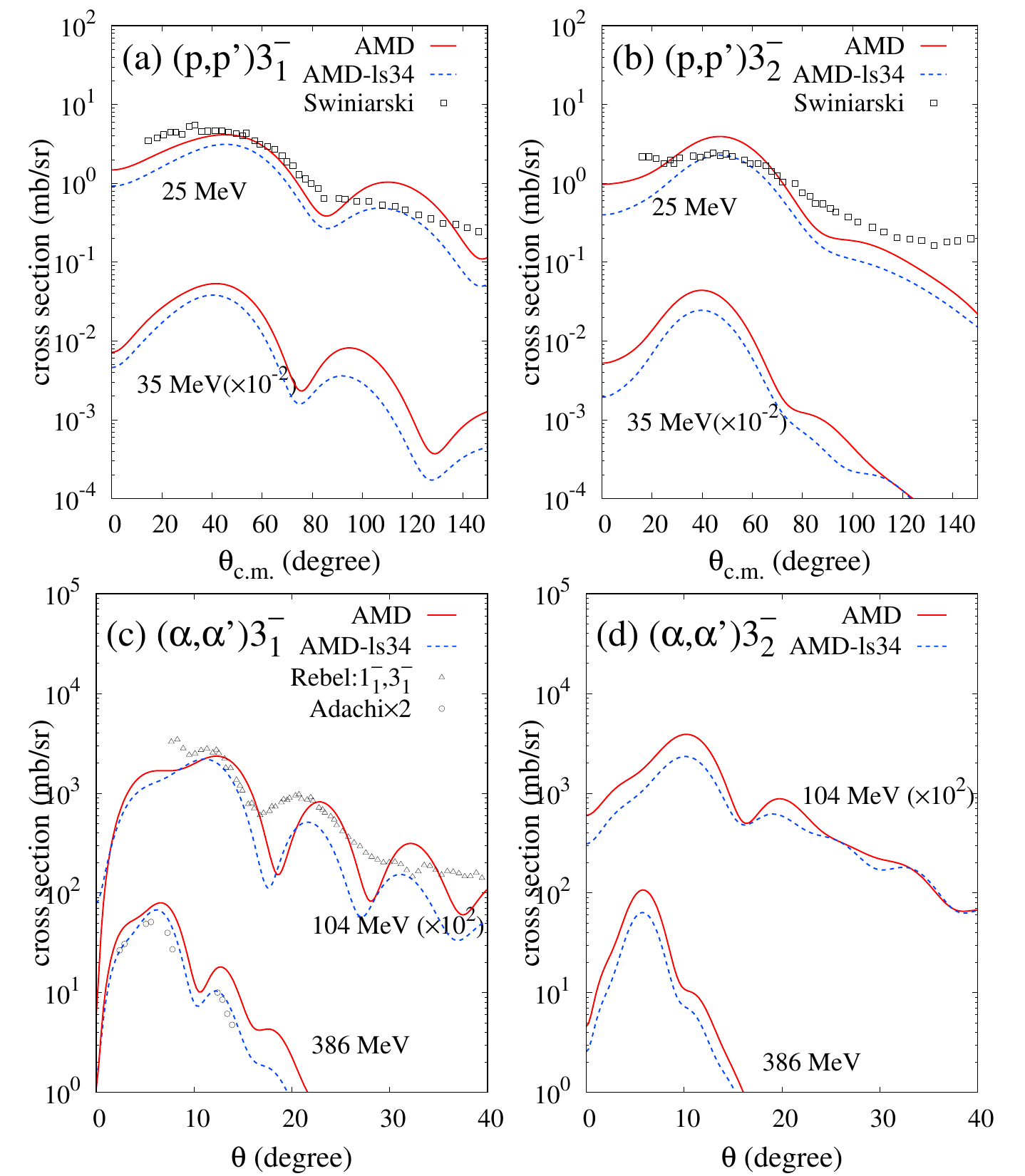}
  \caption{
Cross sections of proton and $\alpha$ inelastic scattering calculated with 
MCC using the  AMD (default) 
and AMD-ls34 (modified spin-orbit strength) densities, and the experimental cross sections.
The $(p,p')$ cross sections at $E_p=25$ and 35 MeV to the 
(a) $3^-_1$ and 
(b) $3^-_2$ states and the $(\alpha,\alpha')$ cross sections 
at $E_\alpha$=146, and 386 MeV
to the (c) $3^-_1$ and (d) $3^-_2$ states. . 
The 
$(p,p')$ data at 
$E_p=24.5$~MeV~\cite{deSwiniarski:1976rcl}
and $(\alpha,\alpha')$ data 
at $E_\alpha=$104~MeV~\cite{Rebel:1972nip} and  
386 MeV~\cite{Adachi:2018pql} are also shown. 
The $(\alpha,\alpha')$ data at 386 MeV of Ref.~\cite{Adachi:2018pql} are multiplied by a factor of two.
The $(\alpha,\alpha')$ data 
at $E_\alpha=$104~MeV in the panel (c) are not cross sections for an individual state but 
may contain $1^-_1$ and $3^-_1$ contributions around  5.7~MeV.
  \label{fig:cross-ls}}
\end{figure*}

The mixing of the $K^\pi=2^-$ and $K^\pi=0^-_1$ bands 
also affects the $3^-_1$ and $3^-_2$ cross sections of proton and $\alpha$ inelastic scattering 
via the transition densities. 
In Fig.~\ref{fig:cross-ls}, the $3^-_1$ and $3^-_2$ cross sections 
calculated with AMD and AMD-ls34, and experimental data are compared. 
For the $(p,p')$ cross sections at $E_p=$25 and 35~MeV, the AMD-ls34 calculation obtains 
smaller cross sections of the $3^-_1$ and $3^-_2$ states than the original AMD result. 
In particular, the suppression at the second peak of the $3^-_1$ cross sections is remarkable 
and shows a better agreement with the $3^-_1$ data at $E_p=$25 MeV.
The $3^-_2$ cross sections are suppressed 
in the whole region of angles. As a result, the agreement with the $3^-_2$ data is improved at the first peak 
but gets somewhat worse at the second peak.

For $\alpha$ scattering to the $3^-_1$ state, peak positions shift to forward angles 
in the AMD-ls34 result probing the longer tail of the transition densities, which is caused by 
mixing of the $K^\pi=0^-_1$ cluster component. 
Compared to the experimental $3^-_1$ cross sections at $E_\alpha=386$~MeV, 
a good agreement is obtained by AMD-ls34. This result may support significant mixing of the 
$K^\pi=0^-_1$ cluster component in the $3^-_1$($K^\pi=2^-$) state. For the 
$(\alpha,\alpha')$ cross sections at $E_\alpha=104$~MeV, 
the data observed for 5.7 MeV
are not enough of high quality to discuss detailed features of the $3^-_1$ state because 
they contain large uncertainty from the $1^-_1$ contribution. 
As for the $3^-_2$ cross sections, the AMD-ls34 result predicts smaller cross sections than the 
AMD result because of the destructive mixing of the $K^\pi=2^-$ component in the 
$3^-_2$($K^\pi=0^-_1$) state.

In the present analysis with AMD and AMD-ls34,
we can say that possible mixing of the $K^\pi=0^-_1$ cluster component 
in the $3^-_1$ state can be probed by 
proton and $\alpha$ cross sections through the transition densities. 
The better agreements of the AMD-ls34 result 
with the $(p,p')$ data
at $E_p=25$~MeV and ($\alpha,\alpha')$ data at $E_\alpha=386$~MeV
supports the significant outer tail of the $3^-_1$ transition densities 
and favors the state mixing case.
For the $3^-_2$ state, experimental data are not enough to draw 
an answer to the state mixing in the $3^-_2$ state.

\section{Summary} \label{sec:summary}

The structure and transition properties of the 
$K^\pi=0^+_1$,  $K^\pi=2^-$, and $K^\pi=0^-_1$ bands
of  $^{20}$Ne 
were investigated with the microscopic structure and reaction calculations 
via proton and $\alpha$ scattering off $^{20}$Ne.

In the structure calculation of $^{20}$Ne with AMD, 
$^{16}\textrm{O}+\alpha$ cluster structures were obtained in the parity-doublet 
$K^\pi=0^+_1$ and $K^\pi=0^-_1$ bands, and the $^{12}\textrm{C}+2\alpha$-like structure
was obtained in the  $K^\pi=2^-$ band. The AMD calculation reproduced 
the experimental $B(E2)$ of in-band transitions. It also described the experimental 
form factors of the $0^+_1$,  $2^+_1$, and $4^+_1$ states.

The MCC calculations with the Melbourne $g$-matrix $NN$ interaction
were performed for proton and $\alpha$ scattering off  $^{20}$Ne 
using the AMD densities of $^{20}$Ne. 
The MCC calculations reasonably reproduced the observed cross sections of 
proton scattering at $E_p=25$--35 MeV and $\alpha$ scattering at $E_\alpha=104$--386 MeV. 
Transition properties from the ground to excited states were discussed via the 
reaction analyses of proton and $\alpha$ inelastic processes.

The mixing of the  $K^\pi=2^-$ and  $K^\pi=0^-_1$ bands 
in the $3^-_1$ and $3^-_2$ states
was investigated in the analyses of AMD~(default) with almost no mixing
and AMD-ls34~(a modified spin-orbit strength) with the state mixing.
The former calculation (AMD) significantly 
underestimates the experimental $B(E3;3^-_1\to 0^+_1)$, 
while the latter (AMD-ls34) calculation obtains a better result for $B(E3;3^-_1\to 0^+_1)$ 
because the $E3$ transition strength is enhanced by mixing of the $K^\pi=0^-_1$ cluster component. 
The state mixing of the $3^-_1$($K^\pi=2^-$) and $3^-_2$($K^\pi=0^-_1$) states also 
affects the $E3$ transition densities from the ground state, 
which can be probed by $(p,p')$ and $(\alpha,\alpha')$ cross sections, in principle. 
The detailed analysis of proton and $\alpha$ cross sections for the  $3^-_1$ and $3^-_2$ states
was performed by the MCC calculations with AMD and AMD-ls34. 
The observed $(p,p')$ data
at $E_p=25$~MeV and ($\alpha,\alpha')$ data at $E_\alpha=386$~MeV
seems to support the mixing of the $K^\pi=0^-_1$ cluster component
in the $3^-_1$($K^\pi=2^-$) state. 

It should be commented that 
applicability of the present MCC approach with the Melbourne $g$-matrix $NN$ interaction
for low-energy proton scattering in the $E_p \lesssim 30$~MeV range has not been well examined yet.
In order to clarify the properties of the $3^-_1$ and $3^-_2$ states, 
further detailed data of proton and $\alpha$ scattering at various energies are needed. 

\begin{acknowledgments}
The authors thank Dr.~Kimura and Dr.~Kawabata for fruitful discussions.
The computational calculations of this work were performed by using the
supercomputer in the Yukawa Institute for theoretical physics, Kyoto University. This work was partly supported
by Grants-in-Aid of the Japan Society for the Promotion of Science (Grant Nos. JP18K03617, JP16K05352, and 18H05407) and by the grant for the RCNP joint research project.
\end{acknowledgments}

\end{document}